\documentclass[preprint,12pt]{elsarticle}

\usepackage{graphicx}        % standard LaTeX graphics tool

\usepackage{amssymb}

\journal{Applied Mathematics Letters}

\newproof{prf}{Proof}

\begin{document}

\begin{frontmatter}



\title{Chaos of the logistic equation with piecewise constant argument} 


\author[address1]{M. Akhmet\corref{cor1}}
\ead{marat@metu.edu.tr}
\author[address1,address2]{D.~Alt{\i}ntan}
\ead{altintan@metu.edu.tr}
\author[address3]{T. Ergen\c{c}}
\ead{tergenc@atilim.edu.tr}
\address[address1]{Institute of Applied Mathematics, Middle East Technical
University,\\ 06531 Ankara, Turkey}
\address[address2]{Sel\c{c}uk University, 42697 Konya, Turkey}
\address[address3]{At{\i}l{\i}m University, 06836, Ankara, Turkey}
\cortext[cor1]{Corresponding author}

%\address{}

\begin{abstract}
We consider  the logistic equation with  different types of the piecewise 
constant argument. It is proved that  the equation generates chaos and intermittency.
Li-Yorke chaos is obtained as well as the chaos through period-doubling  route.
Basic plots are presented to show  the complexity  of   the behavior.   
\end{abstract}


\end{frontmatter}


%\section{}
%\label{}

\section{Introduction} \label{int}

The first papers about simple population models with
complex dynamics are  \cite{mo:76,may:76}.  
The main method of analysis of these models is  the reduction to discrete equations:
the logistic and  Ricker's 
equation \cite{ric:54}.
  
In \cite{lg:99}, Liu and Gopalsamy investigated the following  equation with 
piecewise constant  argument
\begin{equation}
\label{e5}
\frac{d\,x(t)}{dt}= r\,x(t)\,\Big\{1-a\,x(t)-b\,x([t]) \Big\},\:\:\:t>0,   
\end{equation}
where \(r, a, b\) are positive constants and 
\([.]\) denotes the greatest integer function.
The authors showed that for certain parameter 
   values of \(a\) and \(b\), equation (\ref{e5})  generates Li-Yorke chaos
\cite{ly:75}.
   
In the present  paper  the approaches  of  \cite{mo:76,may:76} and  \cite{lg:99} 
are developed 
   for different types of the logistic equation with piecewise constant argument. 
   Transformations of the dependent and independent variables are used to obtain
   convenient discrete equations for the dynamic analysis. The Li-Yorke theorem 
\cite{ly:75} is referred to prove chaos. 
   A connection between solutions  of continuous models and discrete equations is
used to make appropriate simulations. 
   
In the paper following equations are considered:
\begin{eqnarray}
\label{e7}\frac{d\,x(t)}{dt}&=&(a-b\,x([t]))\,x([t]),\\
\label{e8}\frac{d\,x(t)}{dt}&=&(a-b\,x([t+1]))\,x(t),\\
\label{e9}\frac{d\,x(t)}{dt}&=&(a-b\,x([t+1]))\,x([t+1]).
\end{eqnarray}  
It is  seen that  we suggest to involve not only delayed, but also advanced
arguments  in 
    the population models. Although the role of delay in the population dynamics
    has been discussed vitally \cite{mur:03}, the anticipation phenomena has not 
    been considered yet. Anticipatory assumption 
    in a population model may mean that \textit{a will, a wish, an anticipation}
    is taken into account. It can be assumed that anticipation is 
    a prediction reached by the decisions of the present time. We introduce 
    anticipation in our population models via function  \([t+1]\) \cite{aopw:06}.
    
We show that for critical values of the parameters \(a, b\) the 
    solutions of the differential equations (\ref{e7}),(\ref{e8}),(\ref{e9})
    show intermittency which is ``almost periodic'' behavior 
    interrupted by chaotic motions \cite{pm:80,hhs:82}.     

In  the  population models, we consider \(x(t)=N(t)-N_0\),
  where \(N(t)\) denotes the size of the population at time 
  \(t\) and \(N_0\) is a positive integer, let say,  the average value of a population.
  Thus,  \(x\) does not represent the size of the population and it can be 
negative.

\section{Analysis of the equations}\label{secnum1}
Let us start with equation (\ref{e7}). If \(t\in[k,k+1)\), \(k \in \mathbb{N}_{0}\),
it   takes the following form 
\begin{equation}
\label{e11}
\frac{d\,x(t)}{dt}=(a-b\,x_{k})\,x_{k},    
\end{equation}
then, 
\begin{equation}
\label{e12}
x(t)=x_{k}+(a-b\,x_{k})\,x_{k}\,(t-k),    
\end{equation}
and, hence
\begin{equation}
\label{e13}
x_{k+1}=x_{k}\,(1+a-b\,x_{k}).    
\end{equation}
If one makes the change of variable \(q_{k}=\frac{b}{1+a}\,x_{k}\)
in  (\ref{e13}), then obtains 
\begin{equation}
\label{e14}
q_{k+1}=\mu\,q_{k}\,(1-q_{k}),    
\end{equation}
where \(\mu=1+a\).
The right-hand side of equation (\ref{e14})
is the logistic map 
\begin{equation}
\label{e15}
G(q)=\mu\,q\,(1-q).    
\end{equation}
When \(\mu\cong 3.57\), \(G\) generates chaos through period-doubling 
 (see \cite{str:94} for more details). Since this map is obtained 
 from the solution of the equation (\ref{e11}) and \(\mu=1+a\), 
 it is obvious that equation (\ref{e7}) can generate chaos for 
 \(a\cong 2.57\).

In \cite{str:94}, one can find that \(G\) has intermittent
    behavior at \(\mu=3.8282\). Therefore, equation (\ref{e12}), and consequently,
(\ref{e7}),
    displays  intermittency, too.
   
\begin{figure}[h]
  \centering  
  \includegraphics[width=3.75in]{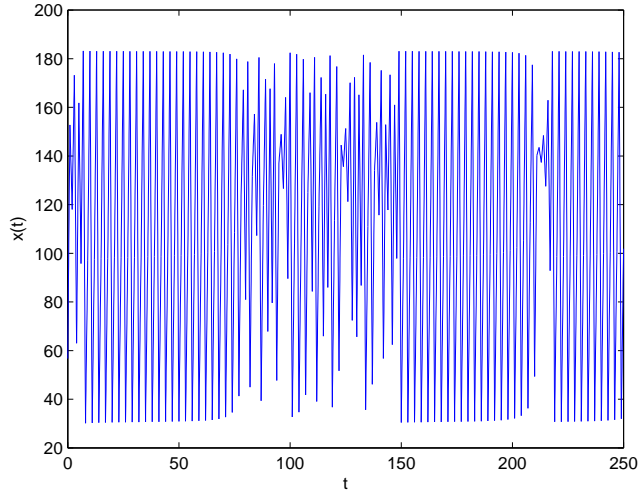}                
  \caption{The solution of the differential equation
\(\frac{d\,x(t)}{dt}=(a-b\,x([t]))\,x([t])\) with 
  \(x(0)=56.7148\), \(a=2.8282\), \(b=1/50\).}
  \label{fig:intermittency x}
\end{figure}

Let \(a=2.8282 \), \(b=1/50\) and \(x(0)=56.7148\). One can see the intermittency
phenomena 
     for equation (\ref{e7}) in Figure \ref{fig:intermittency x}.

Let us consider another equation (\ref{e8}). If \(t\in [k,k+1)\), then
\begin{equation}
\label{e110}
\frac{d\,x(t)}{dt}=(a-b\,x_{k+1})\,x(t),    
\end{equation}
and 
\begin{equation}
\label{e160}
x(t)=x_{k}\,e^{\int_{k}^{t} (a-b\,x_{k+1})\,ds}.
\end{equation}
Hence, 
\begin{equation}
\label{e17}
x_{k}=x_{k+1}\,e^{(b\,x_{k+1}-a)}\:\:\: (k \in \mathbb{N}_{0}).
\end{equation}
Now the transformation \(k=-n\), \(r_{n}=-x_{1-n}\), in (\ref{e17}) yields 
\begin{equation}
\label{e18}
r_{n+1}=r_{n}\,e^{-b\,r_{n}-a},
\end{equation}
where \(n\) is a negative integer. The right-hand side of equation 
(\ref{e18}) is a function of the form 
  \[T(r)=r\,e^{-a-b\,r}.\]

In their article, May and Oster \cite{mo:76} discussed the behavior of 
   the following discrete-time equation:
\begin{equation}
\label{mo}
X_{t+1}=X_{t}\,e^{\gamma\,(1-X_{t})},\:\:\: \gamma>0. 
\end{equation} 
   They proved that for certain values of \(\gamma\) equation 
   (\ref{mo}) has fixed points of period \(k\) for \(k=1,2,3,\ldots\)
   and it generates chaos. Below we will try to extend their results to equation
(\ref{e160})
   for \(a=-\gamma\) and \(b=0.01\).

\begin{figure}[h]
  \centering   
  \includegraphics[width=3.75in]{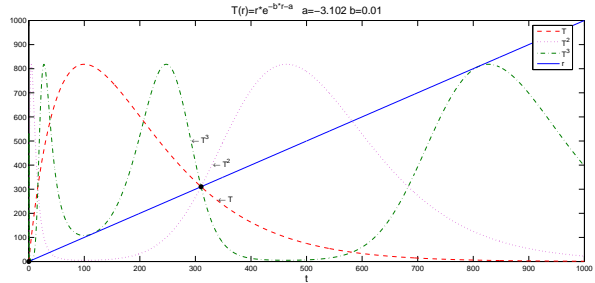}                
  \caption{The graphs of \(T(r),\,T^{2}(r),\,T^{3}(r)\) and \(y=r\) when
\(a=-3.102\) and \(b=0.01\).}
  \label{periodic2teget}
\end{figure}

\begin{figure}[hb]
  \centering 
  \includegraphics[width=3.75in]{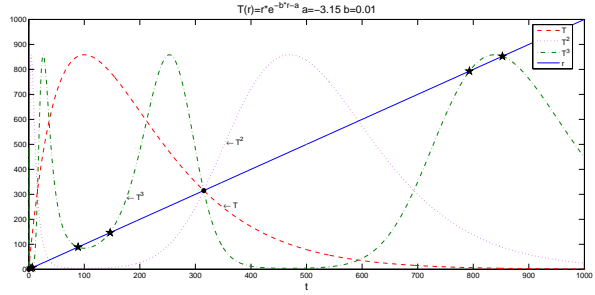}                
  \caption{The graphs of \(T(r),\,T^{2}(r),\,T^{3}(r)\) and \(y=r\) when \(a=-3.15\)
and \(b=0.01\).}
  \label{periodic3}  
\end{figure}

Now let us consider the fixed points of \(T\), \(T^{2}\) and \(T^{3}\) 
    for different values of \(a\) with \(b=0.01\). Consider the value 
\(a_{c}=-3.102\) which is borrowed from \cite{mo:76}, the mapping \(T^{3}\) is
tangent to \(y=r\) line, as shown in Figure \ref{periodic2teget}. 
When \(a=-3.15\), the mapping \(T^{3}\) has extra fixed points 
   which are denoted by black stars in Figure \ref{periodic3}. 
   Then, there exist period three points which are not period one and  two. 
   Consequently, equation (\ref{e8}) admits the chaos through 
   Li and Yorke theorem \cite{ly:75}.   

\begin{figure}[h]
  \centering  
  \includegraphics[width=3.75in]{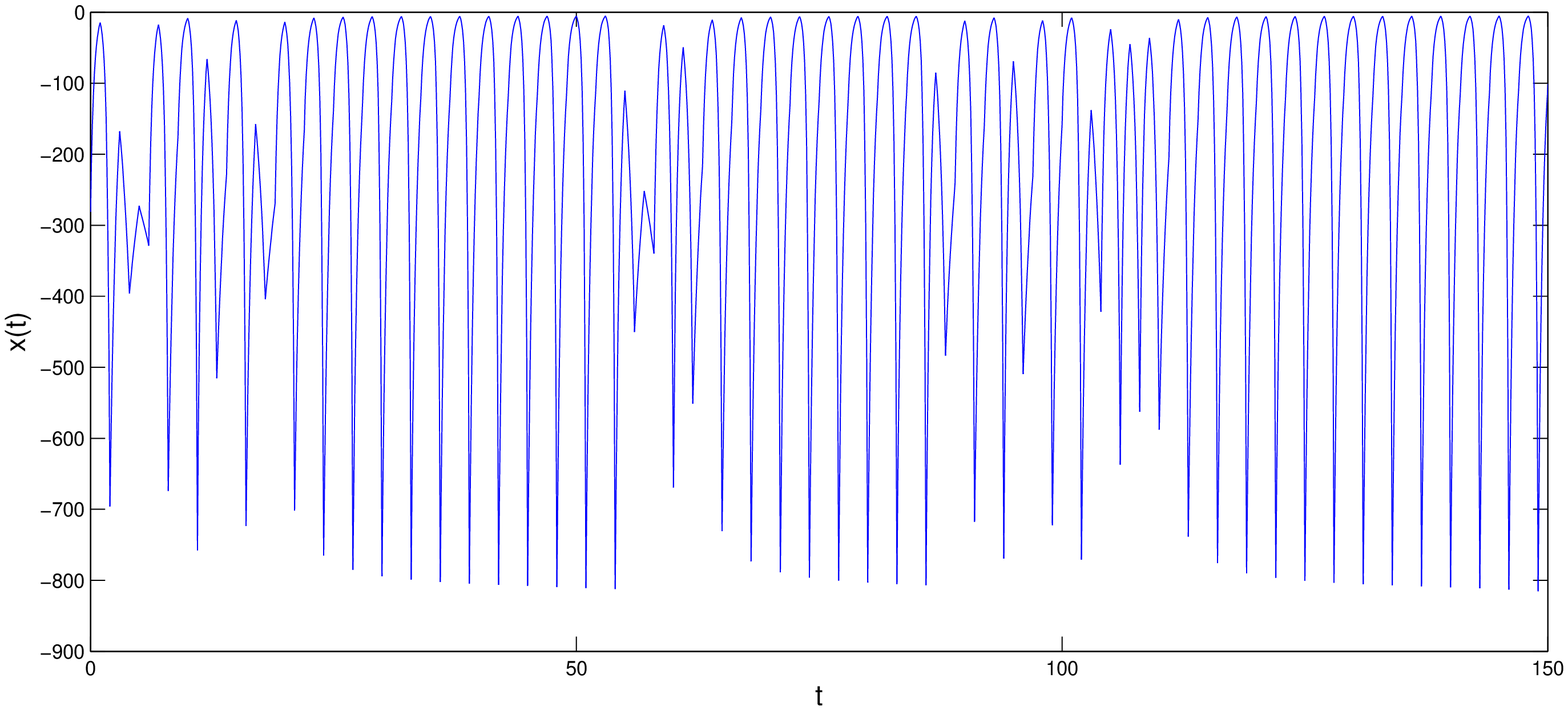}                
  \caption{The graph of \(x(t)\)  for \(a=-3.1\), \(b=0.01\), \(x_{151}=-5\).}
   \label{fig:intermittency x(t)}
\end{figure}

For  values  of \(a\) just above \(a_{c}\), the system displays intermittency.
   In common, simulations of the corresponding discrete equation (\ref{e18})
   are realized, but we propose to  see the complex behavior 
   in its original form. Thus, to  compute \(x(t)\)  for  \(t\geq 0\) let  us apply 
the following program. First, we 
   fix  \(r_{n}\)  with a negative integer \(n\). 
   Then, we calculate  the sequence
\(\left\{r_{n+1},r_{n+2},\ldots,r_{1},r_{0}\right\}\), by  using (\ref{e18}) and,
then \(k=-n\), \(r_{n}=-x_{1-n}\) and \(x_{0}=x_{1}\,e^{(b\,x_{1}-a)}\).
   Substituting  values of \(x_{k}\) and \(x_{k+1}\) in  (\ref{e160}), we  obtain the 
   solution of equation (\ref{e110}). When \(a=-3.1\), \(b=0.01\) and
\(r_{-150}=5\), the result  of simulation is seen in the Figure
\ref{fig:intermittency x(t)}.

\begin{figure}[h]
  \centering    
  \includegraphics[width=3.75in]{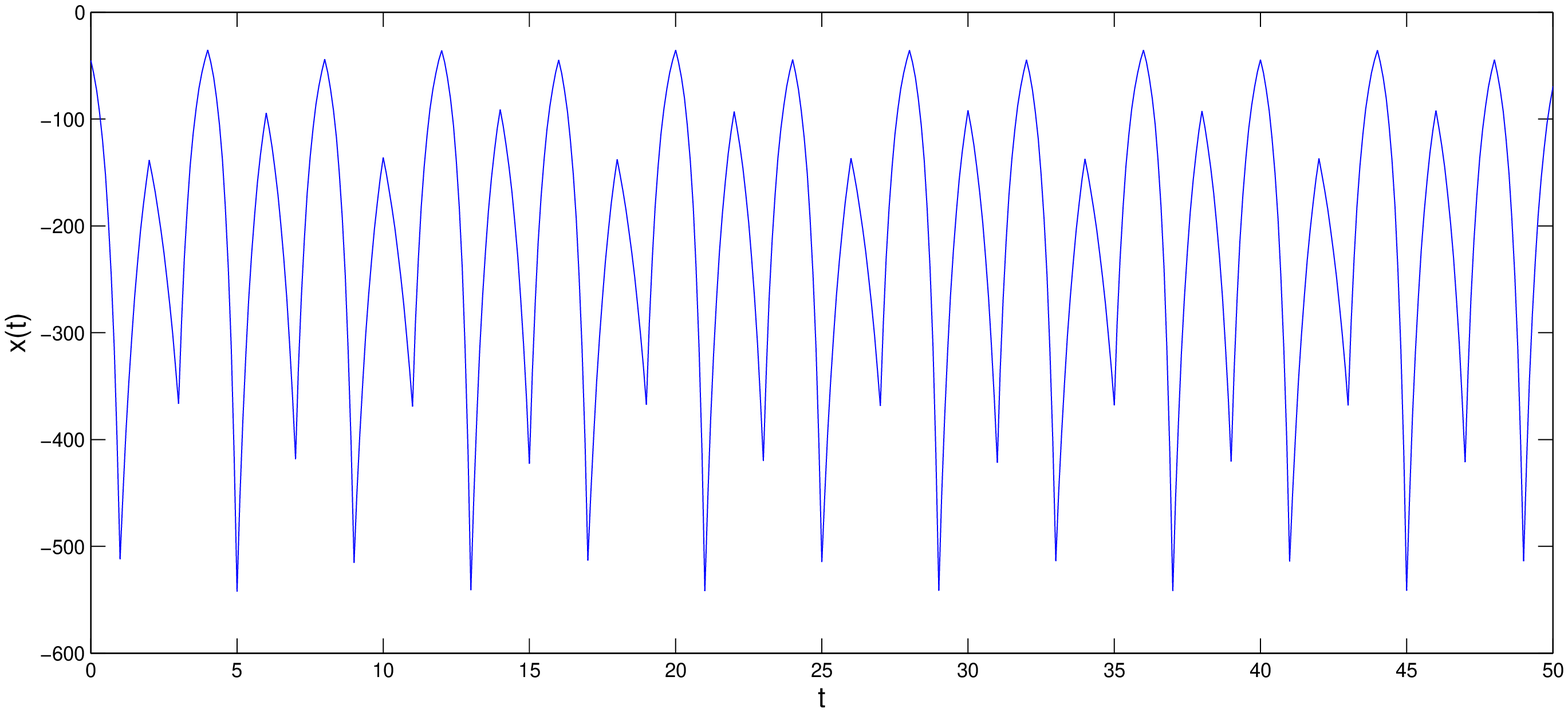}                
  \caption{The graph of \(x(t)\) for \(a_{\infty}=-2.6924...\) and
\(b=0.01,\:x_{51}=-70.2344\).}
  \label{chaos_2}
\end{figure}

Let \(a_{\infty}=-2.6924...\) and \(b=0.01,\:r_{-50}=70.2344\). In  Figure
\ref{chaos_2}, one can 
    see that equation (\ref{e160}) generates chaos.
    
Our last system is as follows 
\begin{equation}
\label{e19}
\frac{d\,x(t)}{dt}=(a-b\,x([t+1]))\,x([t+1]),    
\end{equation}
where \(a\) and  \(b\) are constants with \(a\neq 1.\)
The corresponding discrete-time equation is
\begin{equation}
\label{e20}
x_{k}=x_{k+1}(1-a+b\,x_{k+1}),\:\:\:(k\in \mathbb{N}_{0}).    
\end{equation}
Let \(k=-n\) and \(r_{n}=\frac{b}{a-1}\,x_{1-n}\), the last equation 
can be written in the following form 
\begin{equation}
\label{e21}
r_{n+1}=(1-a)\,r_{n}\,(1-r_ {n}),    
\end{equation}
where \(n\) is a negative integer. The right-hand side of equation (\ref{e21}) is a
function of the form
$P(r)=\mu \,r\,(1-r),$
where \(\mu=1-a\). Similarly  to  the  equation (\ref{e15}),  the last  one  
generates 
complex dynamics. 
 
\section{Conclusion}\label{con}   

We discuss the  complex behavior of different types of 
logistic equations with piecewise constant argument of delay and advance types. 
The idea of anticipation and piecewise constant argument are used together.       
Transformations of the space and time variables are used to obtain  proper
discrete-time equations. 
The parameter values of the discrete-time equations which cause chaos and intemittency 
are utilized to get analogues for continuous solutions. Simulations of the
continuous dynamics are given.

%\bibliographystyle{elsarticle-num}   
%\bibliography{altintan_akhmet}




%\bibitem{}

\end{document}